\documentclass[preprint,showkeys,aps]{revtex4-2} 
\usepackage{amsmath}
\usepackage{latexsym}
\usepackage{amssymb}
\usepackage{graphics,epstopdf}
\usepackage{amsmath,amssymb,amsthm}
\usepackage{graphicx}
\usepackage{mathrsfs}
\usepackage[colorlinks=true, citecolor=blue, urlcolor=blue ]{hyperref}
\usepackage{cleveref}
\DeclareMathOperator{\cosech}{cosech}
\begin{document}
\title{Information geometry and Gaussian entanglement of thermal states for a charged anisotropic oscillator}
\author{Pinaki Patra}
\thanks{Corresponding author}
\email{monk.ju@gmail.com}
\affiliation{Department of Physics, Brahmananda Keshab Chandra College, 111/2 B. T. Road, Kolkata, India-700108}
\keywords{Information geometry; Gaussian entanglement;  Fisher-Rao metric; Curvature singularities; Charged anisotropic oscillator}
\date{\today}
\begin{abstract}
We investigate the information geometry and Gaussian entanglement of a two-dimensional charged anisotropic harmonic oscillator subjected to a homogeneous magnetic field. Using the exact thermal covariance matrix, we construct the classical Fisher-Rao metric associated with Gaussian Wigner distributions on a three-parameter statistical manifold characterized by inverse temperature, oscillator anisotropy and cyclotron frequency. Closed-form expressions for the intrinsic Gaussian curvature are obtained for three physically relevant submanifolds. In the absence of magnetic coupling, the thermal manifold admits an isometric embedding into three-dimensional Euclidean space and exhibits strictly negative curvature. The ground-state manifold develops a curvature singularity at the isotropic zero-field point, originating from metric degeneracy rather than a singularity of the underlying quantum state.   Employing the Gaussian positive partial transpose criterion, we derive an exact separability condition for the canonical  bipartition and an analytical equation governing the critical temperature for entanglement. Magnetic coupling and oscillator anisotropy are shown to be individually necessary and jointly sufficient for ground-state entanglement. Our results demonstrate that negative curvature, curvature sign reversal and curvature divergence need not signify quantum entanglement, thereby highlighting the distinction between statistical distinguishability and quantum correlations in Gaussian systems.
\end{abstract}
\maketitle
\section{Introduction}
Information geometry is a multidisciplinary branch of statistics used for analysing the space  ($\mathcal{S}=\{\mathcal{P}(x,\vartheta) \vert \vartheta = (\vartheta_1, .., \vartheta_l) \in \Theta \subseteq\mathbb{R}^l \}$) of parametric probability distributions ($\mathcal{P}(\vartheta)$) by introducing a differentiable manifold $\mathcal{M}(\Theta,\mathcal{G}(\vartheta))$, endowed with a Riemannian metric, the Fisher-Rao (FR) metric, given by $ds_{\text{FR}}^2 = \sum_{\mu,\nu} g_{\mu\nu} d\vartheta^{\mu} \otimes d\vartheta^{\nu}$, where the components of metric tensors are defined by
$ g_{\mu\nu}(\vartheta)= \int_{x\subseteq \mathbb{R}^d} dx \mathcal{P}\partial_{\mu} \ln\mathcal{P} \partial_{\nu} \ln\mathcal{P}$ \cite{FR1,FR2,FR3,FR4,FR5,FR6}. Here, $d$ is the phase-space dimension and the notation $\partial_{j}$ stands for partial derivative with respect to $\vartheta_j$. Its geometric properties, including geodesic distances, curvature and metric degeneracies, provide valuable insights into the underlying statistical structure of physical systems. For instance, the FR metric quantifies the infinitesimal distinguishability between nearby densities in a statistical model \cite{Raodistance1,Raodistance2,Raodistance3}. Information-geometric methods have found applications in diverse range of fields including equilibrium statistical mechanics, critical phenomena, parameter estimation, signal processing, black-hole thermodynamics, quantum information theory, and machine
learning \cite{Appl1,Appl2,Appl3,Appl4,Appl5,Appl6,Appl7,Appl8,Appl9,Appl10,Appl11,Appl12,Appl13,Appl14}. In quantum theory, the Fisher information  metric measures how distinguishable nearby quantum states are under small changes to a parameter \cite{quantdis1,quantdis2}. In the present paper, when we refer distinguishability, we set aside debates on the ontology of quantum states \cite{ontology1,ontology2,ontology3,ontology4}. In metrology, the FR information is utilized as a universal tool to probe the resourcefulness of quantum states \cite{quantdis3}. \\
A quantum system under any potential with a local minima, can be approximately described by a harmonic oscillator around the local minima of the potential. Thus, at sufficiently low temperature (for instance below $4$ Kelvin) Gaussian states (GS) play a crucial role for an approximate description of a large class of quantum systems \cite{GS1}. A GS $\hat{\rho}$ can be completely characterized by its first moment vector $\langle X\rangle$ of the phase-space operators $X=(\hat{x}_1,...,\hat{x}_d,\hat{p}_1,...,\hat{p}_d)$, and the covariance matrix (CVM) $\Sigma$, defined by  $\Sigma_{ij}=\text{Tr}(\frac{1}{2}\{X_i,X_j\}-\langle X_i\rangle \langle X_j\rangle)$, where the expectation value of an observable $\hat{\mathcal{O}}$ in the state $\hat{\rho}$ is defined by $\langle \hat{\mathcal{O}}\rangle= \text{Tr}(\hat{\mathcal{O}}\hat{\rho})$ \cite{GS2}. This simplicity places GS in a distinguished position for the characterization of continous variable systems in phase-space description. If the moment vector vanishes (in principle, can always be done by suitable co-ordinate choice), then FR metric for GS can alternatively be defined as $g_{ij}= \frac{1}{2} \text{Tr}[\Sigma^{-1}(\partial_{\vartheta_i} \Sigma) \Sigma^{-1}(\partial_{\vartheta_j} \Sigma)]$, which is known as classical FR metric associated with Gaussian Wigner distributions. The FR distance between two GS, corresponding to CVM $\Sigma_1$ and $\Sigma_2$ is defined as $d_{FR}(\Sigma_1,\Sigma_2)=\sqrt{\frac{1}{2}\sum_{j=1}^{2n}\log^2\lambda_j(\Sigma_1^{-1}\Sigma_2)}$, where $\lambda_j(\Sigma_1^{-1}\Sigma_2)$ are the $j^{th}$ largest generalized eigenvalues, which are solutions of $\mbox{Det}(\Sigma_1-\lambda\Sigma_2)=0$ \cite{FR7}.  $d_{FR}$ is invariant under any congruent transformation $\tilde{\Sigma}=\Upsilon\Sigma \Upsilon^T$, where $\Upsilon\in GL(2n,\mathbb{R})$ is an invertible $2n\times 2n$ matrix over $\mathbb{R}$ \cite{FR8}. Another aspect of GS is that, we have a sufficient condition for the separability of bipartite (say A and B) two mode GS through positive partial transpose (PPT) criteria, which states that the Robertson-Schr\"{o}dinger uncertainty principle (RSUP) $\Sigma+\frac{i\hbar}{2}J_4\ge 0$ will be manifested by $\tilde{\Sigma}+\frac{i\hbar}{2}J_4\ge 0$ for a bonafide CVM, even after a partial transposition ($\Lambda_B$: changes sign of momentum coordinates of the subsystem B and $\Lambda_B:\Sigma\mapsto\tilde{\Sigma}=\Lambda_B\Sigma\Lambda_B^T$) with respect to a subsystem (say B) \cite{PPT1}. Here $\hbar$ is the Planck constant and $J_{2n}$ is the symplectic matrix, which encodes the canonical commutation relations $[\hat{X}_{ j}, \hat{X}_{ k}]=i\hbar (J_{2n})_{jk}$. \\
Since both the FR geometry and PPT-separability are completely characterized by CVM, it is worth studying the aspects of their complementarity. In particular, since the FR-geometry characterizes the local statistical geometry of a parametrized family of states, whereas entanglement depends on the subsystem decomposition and the quantum correlations between its constituents, establishing the extent to which these two structures are related is therefore of considerable interest.
In this paper, we explore the interconnection between FR-geometry and PPT-separability for a  system of charged anisotropic oscillator confined to a homogeneous magnetic field. The system is in thermal equilibrium with a heat bath, with inverse temperature $\beta$. The magnetic interaction introduces  coupling between the phase-space coordinates, while the anisotropy breaks rotational symmetry of the Hamiltonian $\hat{H}_B$. The interplay between these two mechanisms  generates intermode correlations and influences the statistical distinguishability of the corresponding thermal states $\hat{\rho}_{\beta} = \frac{e^{-\beta\hat{H}_B}}{\mathscr{Z}_{\beta}}$, where the canonical partition function is $ \mathscr{Z}_{\beta} = \text{Tr}(e^{-\beta\hat{H}_B})$ \cite{GT1,GT2,GT3,GT4}.  The resulting family of GS provides a suitable analytical model for investigating the combined influence of anisotropy, magnetic coupling and temperature on information geometry and quantum entanglement. \\
 \\
We demonstrate that the thermal information manifold in the absence of magnetic coupling admits an isometric embedding into three-dimensional Euclidean space and possesses strictly negative Gaussian curvature at finite temperatures. The ground-state manifold exhibits a curvature singularity associated with metric degeneracy at the isotropic zero-field point. In particular, the isotropic and zero-field limits of the Gaussian curvature are shown to be noncommutative. For the isotropic thermal oscillator, the Gaussian curvature can change sign with temperature, These results illustrate how the restoration of symmetry and the reduction of the number of independent statistical parameters can lead to nonuniform limiting geometries. Furthermore, we derive an exact analytical separability criterion for the canonical Cartesian bipartition of the oscillator. We establish that oscillator anisotropy and magnetic coupling are individually necessary and jointly sufficient for ground-state entanglement. A comparison between the information-geometric and separability properties reveals that negative curvature, curvature sign reversal and even curvature divergence can occur in the absence of quantum entanglement. The present analysis thus provides an explicit example demonstrating that intrinsic FR curvature cannot, solely through its sign or singular behaviour, serve as a universal indicator of Gaussian entanglement.\\
The organization of the paper is as follows. In section II, we derive the exact thermal covariance matrix and construct the corresponding  FR metric.  Section III is devoted to the information geometry of the oscillator in the absence of an external magnetic field. In section IV, we investigate the ground-state manifold, with particular emphasis on curvature singularities and the noncommutativity of geometric limits. The information geometry of the isotropic oscillator is studied in section V. In section VI, we employ the Gaussian PPT criterion to investigate separability and discuss its relationship with the intrinsic curvature of the FR manifold. Finally, our principal findings and their implications are summarized in section VII.
\section{Information geometry of an anisotropic oscillator in a homogeneous external magnetic field }
If a two-dimensional charged anisotropic harmonic oscillator of mass $m$ and frequency $\omega_{xy}=(\omega_x,\omega_y)$ is confined to a plane perpendicular to a homogeneous magnetic field $\vec{B}=B\hat{z}$, then in the symmetric gauge (vector potential $\vec{A}=\frac{1}{2}B(-y,x)$) the Hamiltonian reads
\begin{equation}\label{HB}
	\hat{H}_B= \frac{1}{2m}(\hat{p}_x^2 +\hat{p}_y^2) + \frac{1}{2}m (\omega_1^2 x^2+ \omega_2^2 y^2) -\frac{1}{2}\omega_c \hat{L}_z,
\end{equation}
where the cyclotron frequency $\omega_c$, the effective frequency $\omega_{\text{ef}}=\text{Diag}(\omega_1,\omega_2)$  are given by
\begin{equation}
	\omega_c=\frac{qB}{mc},\; \omega_1^2=\omega_x^2 +\frac{1}{4}\omega_c^2,\; \omega_2^2=\omega_y^2 +\frac{1}{4}\omega_c^2.
\end{equation}
The $z$-component of orbital angular momentum is $L_z=xp_y-yp_x$. Here, $q$ is the charge of the particle and $c$ is the speed of light in vacuum. If we arrange the phase-space operators as $X=( \hat{X}_{1}, \hat{X}_{2}, \hat{X}_{3},\hat{X}_4)^T=(\hat{x},\hat{y},\hat{p}_x,\hat{p}_y)^T$, then ~\eqref{HB} can be written in the following quadratic form
\begin{equation}\label{HBquadratic}
	\hat{H}_B=\frac{1}{2} X^T \mathcal{H} X,\; \mbox{with}\; 
	\mathcal{H}= \left( \begin{array}{cc}
	m\omega^2 & -\frac{\omega_c}{2}J_2 \\
		\frac{\omega_c}{2} J_2 & \frac{1}{m} \mathbb{I}_2
	\end{array}\right),\;
J_{2n}=\left(\begin{array}{cc}
0 & \mathbb{I}_{n}\\
-\mathbb{I}_{n} & 0
\end{array}\right),
\end{equation}
where $\mathbb{I}_{n}$ denotes the $n\times n$ identity matrix.
The superscript $T$ denotes matrix transposition.  The canonical commutation relations are given by
 \begin{equation}
 	[\hat{X}_{ j}, \hat{X}_{ k}]=i\hbar (J_4)_{jk}; j,k\in \{1,2,3,4\},
 \end{equation}
 where $(J_4)_{jk}$ is the $(j,k)$th element of the symplectic matrix $J_{4}$, and $\hbar$ is the Planck's constant. 
By performing a symplectic transformation, the quadratic Hamiltonian ~\eqref{HBquadratic} can be diagonalized in terms of annihilation ($\hat{a}_{1,2}$) and creation ($\hat{a}_{1,2}^\dagger$) operators as follows.
\begin{equation}\label{HN}
	\hat{H}_B =\hbar\sum_{j=1}^{2} \Omega_j (\hat{n}_j+1/2),\; \mbox{with}\; \hat{n}_j = \hat{a}_j^\dagger \hat{a}_j,
\end{equation}
where the two normal-mode frequencies $\Omega_j\; (j=1,2)$ are given by
\begin{equation}
\Omega_{1,2}^2=	\omega_{\pm}^2 = \frac{1}{2}[ (\omega_1^2+\omega_2^2 + \frac{1}{2}\omega_c^2) \pm \sqrt{(\omega_1^2 -\omega_2^2)^2 + 2\omega_c^2 (\omega_1^2 + \omega_2^2)}].
\end{equation}
Note that the normal-mode operators ($\hat{a}_{1,2}, \hat{a}_{1,2}^\dagger$) satisfy the usual bosonic commutation relations $[\hat{a}_j,\hat{a}_k^\dagger]=\delta_{jk}$. The energy eigenvalues of Eq.~\eqref{HN} can thus be written as 
\begin{equation}\label{En1n2}
	E_{n_1,n_2}= \hbar [(n_1+1/2)\Omega_1 + (n_2+1/2)\Omega_2],\; \mbox{where}\; n_1,n_2=0,1,2,...
\end{equation}
If the system is in thermal equilibrium with a heat bath of inverse temperature $\beta=\frac{1}{K_B T}$, where $T$ is the absolute temperature, and $K_B$ is the Boltzmann constant, then the thermal density operator $\hat{\rho}_{\beta}$ and the canonical partition function $\mathscr{Z}_{\beta}$ are defined by
\begin{equation}\label{densitypartition}
	\hat{\rho}_{\beta}= \frac{e^{-\beta\hat{H}_B}}{\mathscr{Z}_{\beta}},\; \mathscr{Z}_{\beta}= \text{Tr}(e^{-\beta\hat{H}_B}).
\end{equation}
Using the energy eigenvalues~\eqref{En1n2} in ~\eqref{densitypartition}, we obtain
\begin{equation}
\mathscr{Z}_{\beta}= (4\sinh\nu_1 \sinh\nu_2)^{-1},\; \mbox{with}\; \nu_j= \hbar\beta\Omega_j/2.
\end{equation}
The expectation value of an observable $\hat{\mathcal{O}}$ in the state $\hat{\rho}_{\beta}$ is defined by $\langle \hat{\mathcal{O}}\rangle_{\hat{\rho}_\beta}= \text{Tr}(\hat{\mathcal{O}}\hat{\rho}_\beta)$.
Since $\hat{H}_B$ is invariant under parity ($\mathcal{P}: X\mapsto -X$), $\hat{\rho}_\beta$ is also parity invariant. Hence,
\begin{eqnarray}
	\langle \hat{X}_j\rangle_{\rho_\beta} = \text{Tr}(\hat{X}_j \hat{\rho}_{\beta}) = -\text{Tr}(\hat{X}_j \hat{\rho}_{\beta}) \implies \langle \hat{X}_j\rangle_{\rho_\beta}=0.
\end{eqnarray}
Accordingly, we can define the thermal covariance matrix $\Sigma = (\Sigma_{ij})_{4\times 4}$ by
\begin{equation}
	\Sigma_{ij} =\frac{1}{2} \text{Tr} (\hat{\rho}_{\beta} \{\hat{X}_i, \hat{X}_j\}),\; \mbox{where}\;
	\{\hat{X}_i, \hat{X}_j\}=\hat{X}_i \hat{X}_j+\hat{X}_j \hat{X}_i . 
\end{equation}
Note that the symplectic eigenvalues $\Omega_j$ ($j=1,2$) of $\mathcal{H}$ are obtained from the ordinary eigenvalues of $\mathcal{H}_J=J_4\mathcal{H}$. One can define projectors $\hat{\Pi}_j$ ($j=1,2$) as
\begin{equation}
	\hat{\Pi}_1= -\frac{1}{\Delta_{\Omega^2}}(\hat{\mathcal{H}}_J^2 + \Omega_2^2\hat{\mathbb{I}}),\; 
	\hat{\Pi}_2=\hat{\mathbb{I}}-\hat{\Pi}_1,\;\mbox{where}\; \Delta_{\Omega^2}=\Omega_1^2-\Omega_2^2.
\end{equation}
Using the property that a  matrix satisfies its characteristic polynomial, one can verify that $\hat{\Pi}_j$ are legitimate projectors ($\hat{\Pi}_j^2=\hat{\Pi}_j$). 	The covariance matrix is consequently given by
\begin{equation}\label{sigma1}
	\Sigma = \frac{\hbar}{2} P C,\; \mbox{where} \; C=-\mathcal{H}_J J_4,\; \hat{P}=\sum_{j=1}^{2} \alpha_j \hat{\Pi}_j,\; \mbox{with}\; \alpha_j=\mu_j/\Omega_j,\;\mu_j=\coth\nu_j.
\end{equation}
Henceforth, we adopt units in which $m=1, q=1,c=1$, $\omega_x=1$ and $\omega_y=\delta$. In this choice, we can refer $B$ and $\omega_c$ interchangeably. In particular, we shall consider how the external magnetic field ($B$ or equivalently $\omega_c$), the frequency ratio $\delta$, and the temperature $T$ affect the information geometry of the system. From ~\eqref{sigma1}, we can write the explicit form of $\Sigma$ as
\begin{equation}\label{sigma}
	\Sigma (\beta,\delta,\omega_c)=\frac{\hbar}{2}\left(\begin{array}{cccc}
a_1 & 0 & 0 & c_1 \\
0 & a_2 & c_2 & 0 \\
0 & c_2 & b_1 & 0\\
c_1 & 0 & 0 & b_2
	\end{array}\right),
\end{equation}
where
\begin{eqnarray}
	a_1 &=& \alpha_{\Omega} + (1+\omega_c^2)\alpha_0,\;\; 
		b_1= (1+\omega_c^2/4)\alpha_{\Omega} + (1+ (4+\delta^2)\omega_c^2/4 + \omega_c^4/4)\alpha_0, \label{a1}\\
		a_2 &=& \alpha_{\Omega} + (\delta^2+\omega_c^2)\alpha_0,
		\;\; 
		b_2= (\delta^2+\omega_c^2/4)\alpha_{\Omega}+ (\delta^4+ (1+4\delta^2)\omega_c^2/4 + \omega_c^4/4) \alpha_0, \label{a2}\\
		c_1 &=& - [\alpha_{\Omega} + (1+ 2\delta^2 + \omega_c^2)\alpha_0] \omega_c/2,\;\;
		c_2= [\alpha_{\Omega} + (2+ \delta^2 + \omega_c^2)\alpha_0] \omega_c/2, \label{c1}\\
		&& \mbox{with}\;\;
		 \alpha_{\Omega}= (\Omega_1^2\alpha_2 - \Omega_2^2 \alpha_1)/ \Delta_{\Omega}^2, \; 
		 \alpha_0= (\alpha_1-\alpha_2)/\Delta_{\Omega}^2. \label{c2}
	\end{eqnarray}
Direct calculation yields
\begin{equation}\label{detsigma}
	\text{Det} \Sigma= \hbar^4 \mu_1^2\mu_2^2/16 .
\end{equation}
At absolute zero ($\beta\to\infty; \mu_1=\mu_2=1$), the covariance matrix  reduces to
\begin{eqnarray}
	a_{1}(0) &=& \frac{1+\delta}{w_{+}},\;  \;b_1(0) = \frac{w_{+}}{1+\delta}+ \frac{\omega_c^2(1-\delta)^2}{4\delta(1+\delta)w_{+}},\; 
	c_{1}(0) = \frac{\omega_c(\delta-1)}{2w_{+}},\;
	a_2(0) = \frac{1+\delta}{w_{+}\delta}, \label{covzeroel1}\\
		b_2(0) &=& \frac{w_{+}\delta}{1+\delta}+ \frac{\omega_c^2(1-\delta)^2}{4(1+\delta)w_{+}},\;
		 c_{2}(0)= \frac{\omega_c(\delta-1)}{2w_{+}\delta};\;
	 w_{\pm}=\Omega_1\pm \Omega_2= \sqrt{(1\pm\delta)^2+\omega_c^2}. \label{covzeroel2}
\end{eqnarray}
The argument $(0)$ indicates evaluation at zero temperature. What follows is that, the covariance matrix at absolute zero satisfies the pure-state condition
\begin{equation}
	\Sigma_0 J_4 \Sigma_0=\frac{\hbar^2}{4}J_4, \; \text{where} \; \Sigma_0= \Sigma ( \beta=\infty, \delta, \omega_c).
\end{equation}
 The determinant of $\Sigma_0$ is $\text{Det} \Sigma_0 =\hbar^4/16$, which is consistent with ~\eqref{detsigma}, which reflects that, as the temperature tends to zero, the state becomes a pure Gaussian state, which is consistent with the fact that the oscillator tends to its ground-state as the temperature approaches absolute zero ($T=0$).
Note that in the absence of the magnetic field ($B=0$), the cross-correlation elements $c_1$ (correlation between $\hat{x}-\hat{p}_y$) and $c_2$ (correlation between $\hat{y}-\hat{p}_x$) vanish.  These cross-correlations are therefore induced by the magnetic coupling.
In order to calculate the FR metric corresponding to the covariance matrix~\eqref{sigma}, let us choose the independent parameters $\vartheta = (\vartheta_1, \vartheta_2, \vartheta_3) =(\beta,\delta,\omega_c)$.  The components of the FR metric are given by
\begin{equation}\label{gijdefn}
	g_{ij}= \frac{1}{2} \text{Tr}\left[\Sigma^{-1}\frac{\partial \Sigma}{\partial \vartheta_i} \Sigma^{-1}\frac{\partial \Sigma}{\partial \vartheta_j}\right];\; i,j=1,2,3.
\end{equation}
Throughout this work, the Fisher-Rao  (FR) metric refers to the classical Fisher information metric associated with the Gaussian Wigner distribution of the oscillator state, rather than the symmetric-logarithmic-derivative quantum Fisher information metric.
Substituting Eq.~\eqref{sigma} in Eq.~\eqref{gijdefn}, we can write the metric tensor as
\begin{eqnarray}
	g_{11} &=& \frac{p_1^2+q_1^2+2s_1t_1}{2d_1^2} +  \frac{\bar{p}_1^2+\bar{q}_1^2+2\bar{s}_1\bar{t}_1}{2d_2^2},\\
	g_{22} &=& \frac{p_2^2+q_2^2+2s_2t_2}{2d_1^2}+  \frac{\bar{p}_2^2+\bar{q}_2^2+2\bar{s}_2\bar{t}_2}{2d_2^2},\\
g_{33} &=& \frac{p_3^2+q_3^2+2s_3t_3}{2d_1^2}+  \frac{\bar{p}_3^2+\bar{q}_3^2+2\bar{s}_3\bar{t}_3}{2d_2^2},\\
g_{12} &=& \frac{p_1p_2+q_1q_2+s_1t_2+s_2t_1}{2d_1^2} +  \frac{\bar{p}_1\bar{p}_2+\bar{q}_1\bar{q}_2+\bar{s}_1\bar{t}_2+\bar{s}_2\bar{t}_1}{2d_2^2},\\
g_{13} &=& \frac{p_1p_3+q_1q_3+s_1t_3+s_3t_1}{2d_1^2} +  \frac{\bar{p}_1\bar{p}_3+\bar{q}_1\bar{q}_3+\bar{s}_1\bar{t}_3+\bar{s}_3\bar{t}_1}{2d_2^2},\\
g_{23} &=& \frac{p_2p_3+q_2q_3+s_2t_3+s_3t_2}{2d_1^2} +  \frac{\bar{p}_2\bar{p}_3+\bar{q}_2\bar{q}_3+\bar{s}_2\bar{t}_3+\bar{s}_3\bar{t}_2}{2d_2^2},
\end{eqnarray}
where, for brevity, we define the parameter dependent quantities as
\begin{eqnarray}
p_{i} &=& b_2a_{1,i}-c_1c_{1,i},\; \bar{p}_{i}=b_1a_{2,i} -c_2 c_{2,i},\;
q_{i} = a_1 b_{2,i}-c_1c_{1,i},\; \bar{q}_{i}=a_2 b_{1,i}-c_2c_{2,i},\\
s_i &=& b_2c_{1,i}-c_1b_{2,i},\; \bar{s}_i=b_1c_{2,i}-c_2b_{1,i},\;
t_i=a_1c_{1,i}- c_1a_{1,i},\; \bar{t}_i=a_2c_{2,i}- c_2a_{2,i},\\
d_1 &=& a_1b_2-c_1^2,\; d_2= a_2b_1-c_2^2.
\end{eqnarray}
Here, partial derivatives are denoted by subscripts following a comma. For instance, $a_{1,i}=\frac{\partial a_1}{\partial\vartheta_i}$, etc. The FR line element is given by
\begin{equation}
	ds_{\text{FR}}^2= \sum_{i,j=1}^{3}g_{ij}d\vartheta_{i} d\vartheta_{j}.
\end{equation}
\section{Information geometry in the absence of an external magnetic field} 
At $B=0$ limit ($\omega_c=0$), the normal-mode frequencies are the bare oscillator frequencies $\Omega_{1,2}=\omega_{x,y}$. The covariance matrix  reduces to
\begin{equation}
	\Sigma(\beta,\delta,0)= \frac{\hbar}{2} \text{Diag}(a_{10},a_{20},a_{10},\delta^2 a_{20}),
\text{with}\;	a_{10} = \coth(\hbar\beta/2),
	a_{20} = \delta^{-1}\coth(\hbar\beta\delta/2).
\end{equation}
Accordingly, the metric tensor components for two-parameter family $\vartheta=(\beta,\delta)$  reduce to
\begin{eqnarray}
g_{\beta\beta} &=& \hbar^2 [\cosech^2(\hbar\beta)+\delta^2\cosech^2(\hbar\beta\delta)], \\ g_{\beta\delta }&=& \hbar^2\beta\delta\cosech^2(\hbar\beta\delta),\;
g_{\delta\delta}=\delta^{-2}+ \hbar^2\beta^2\cosech^2(\hbar\beta\delta),
\end{eqnarray}
yielding the line element
\begin{equation}\label{FRmetricB0}
	ds_{FR}^2=\hbar^2 \cosech^2(\hbar\beta)d\beta^2 + \hbar^2 \cosech^2(\hbar\beta\delta)d(\beta\delta)^2 + d(\ln \delta)^2.
\end{equation}
Interestingly, by introducing an embedding into three-dimensional Euclidean space  $(\beta,\delta)\mapsto (u_1,u_2,u_3)$, defined by
\begin{equation}\label{ujdefn}
	u_1=-\ln\mu_1,\; u_2=-\ln\mu_2,\;u_3=\ln\delta,
\end{equation}
the FR-metric ~\eqref{FRmetricB0} can be written as
\begin{equation}
	ds_{FR,3}^2=du_1^2+du_2^2+du_3^2.
\end{equation}
Note that $u_1,u_2,u_3$ are not independent. They define a two-dimensional surface embedded in three-dimensional Euclidean space. In particular, the equation of the surface reads
\begin{equation}\label{u3defn}
	u_3(u_1,u_2)=\ln[\tanh^{-1}(e^{u_2})/\tanh^{-1}(e^{u_1})].
\end{equation}
In other words, the FR manifold (in the absence of an external magnetic field) is  isometric to a surface in Euclidean three-space.\\ The intrinsic Gaussian curvature ($K_G$) for a graph $u_3(u_1,u_2)$ is given by
\begin{equation}\label{KGdefn}
	K_G= \frac{u_{3,u_1u_1} u_{3,u_2u_2}-u_{3,u_1u_2}^2}{(1+u_{3,u_1}^2 + u_{3,u_2}^2)^2}.
\end{equation}
Here we use subscripts following a comma to denote partial derivatives. Using Eq.~\eqref{u3defn} into Eq.~\eqref{KGdefn}, we obtain the exact expression for the Gaussian curvature 
\begin{eqnarray}\label{KGB0eq}
	K_G(B=0)= \frac{-k_{\beta}^2 k_{\beta\delta}^2 }{(1+k_{\beta}^2 + k_{\beta\delta}^2)^2} \left(\frac{\cosh(\hbar\beta)}{k_{\beta}} - 1\right) \left(\frac{\cosh(\hbar\beta\delta)}{k_{\beta\delta}}-1\right), \text{with}\; 
	k_{\theta}:=\frac{\sinh(\hbar\theta)}{\hbar\theta}.
\end{eqnarray}
\begin{figure}
	\includegraphics[width=8cm]{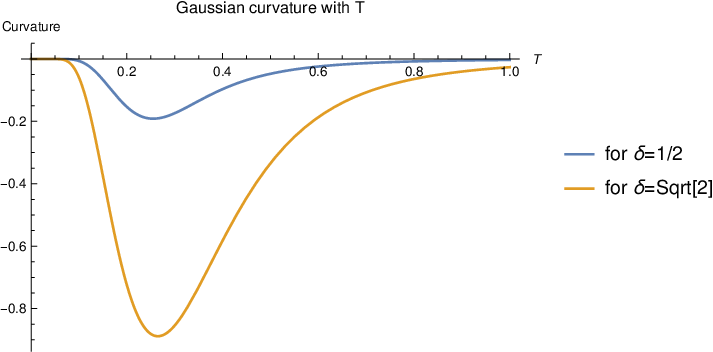}
	\caption{ The Gaussian curvature ($K_G$) is negative at finite temperatures T. The curves correspond to two values of the anisotropy parameter $\delta=1/2,\sqrt{2}$. }
	\label{KGB0}
\end{figure}
Since  $x\coth x>1$ for all  $x>0$, Eq.~\eqref{KGB0eq} implies that $K_G<0$ for all $\beta$ and $\delta$. This behaviour is illustrated in  the figure FIG.~\ref{KGB0}. Thus, in the absence of an external magnetic field, the thermal FR manifold is everywhere negatively curved. In numerical analysis, we have used $\hbar=1$, $K_B=1$, which will be maintained throughout the present study, unless otherwise specified.
\section{Information geometry at absolute zero }
\begin{figure}
	\includegraphics[width=8cm]{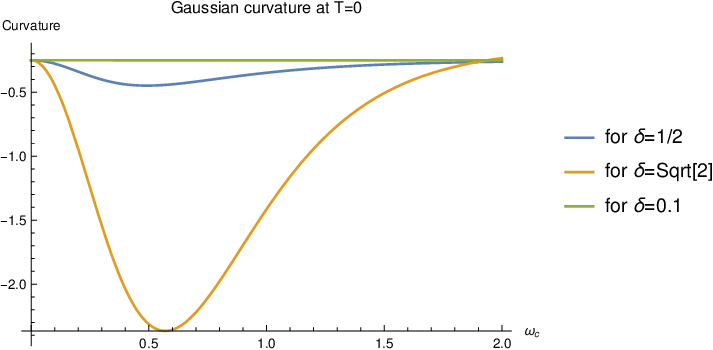}
	\caption{Gaussian curvature $K_G$ as a function of the cyclotron frequency $\omega_c$.  Here we compare variations for three values of the anisotropy parameter $\delta=1/2,\sqrt{2},\delta=0.1$. }
	\label{KGT0}
\end{figure}

At absolute zero ($T=0$), we have the covariance matrix elements ~\eqref{covzeroel1} and ~\eqref{covzeroel2}. Accordingly, the two-parameter family $\vartheta=(\delta,\omega_c)$ yields the FRao line element of the pure-state manifold as
\begin{equation}
ds_{FR}^2 = \left(\frac{1}{\delta^2}-\frac{2\omega_c^2}{\omega_0^4}\right)d\delta^2 + \frac{4\omega_c}{\omega_0^4}(1+\delta)d\delta d\omega_c + \frac{(1+\delta)^2[(1-\delta)^2+\omega_c^2]}{2\omega_0^4\delta}d\omega_c^2,
\end{equation}
from which we obtain the intrinsic Gaussian curvature
\begin{eqnarray}
	K_G(T=0) &=& -\frac{\omega_c^8 + \delta_6\omega_c^6 + \delta_4\omega_c^4 + \delta_2\omega_c^2 + \delta_0}{4[(1-\delta^2)^2+\omega_c^2(2+\omega_c^2)]^2},\label{KG0}\\
\mbox{with}\; \delta_0 &=& (1-\delta^2)^4,\;
	\delta_2= 2(1+\delta)^3(2-6\delta+9\delta^2-3\delta^3),\\
	\delta_4 &=& 2(3+4\delta^3-3\delta^4),\;
	\delta_6= \frac{2(2+2\delta+\delta^2-3\delta^3)}{1+\delta},
\end{eqnarray}
as illustrated in the figure FIG.~\ref{KGT0}.
From Eq.~\eqref{KG0}, we obtain
\begin{equation}\label{lim1}
	\lim\limits_{\omega_c\to 0} K_G(T=0)=-\frac{1}{4} \implies \lim\limits_{\delta\to 1} \lim\limits_{\omega_c\to 0} K_G(T=0)=-\frac{1}{4},
\end{equation}
whereas
\begin{equation}\label{lim2}
	\lim\limits_{\delta\to 1} K_G(T=0)=- \frac{32 + 8 \omega_c^2 + 
	2 \omega_c^4 + \omega_c^6}{4 \omega_c^2 (2 + \omega_c^2)^2} \implies \lim\limits_{\omega_c\to 0} \lim\limits_{\delta\to 1} K_G(T=0)=-\infty.
\end{equation}
In other words, ~\eqref{lim1} and ~\eqref{lim2}  demonstrate that the isotropic and zero-field limits of the Gaussian curvature do not commute. This nonuniform limiting behavior originates from the degeneracy of the FR metric at the isotropic zero-field point.  This nonuniformity can further be manifested through the discontinous nature of $K_G$ at the origin of the parameter space $\epsilon=(\epsilon_\delta,\epsilon_\omega) =(\delta-1,\omega_c)$. Along a path $\epsilon_\omega=\lambda_{\epsilon}\vert \epsilon_\delta\vert^\upsilon$ with $\upsilon>0$, the limiting behavour is path dependent around the origin $(\epsilon_\delta,\epsilon_\omega)=(0,0)$ as
\begin{equation}
	\lim_{\epsilon\to (0,0)} K_G=\left\{\begin{array}{cc}
		-1/4, & \upsilon>2\\
		-1/4-\lambda_{\epsilon}^2/2, & \upsilon=2\\
		-\infty, & 0<\upsilon<2.
\end{array}\right.,
\end{equation}
which demonstrates that the curvature can approach a continuum of finite negative values, as well as diverge, depending on the path of approach. The singular point therefore possesses a highly nonuniform local geometry, even though the underlying Gaussian state remains regular. 
\\
To investigate the intrinsic information geometry of the isotropic oscillator, we shall therefore impose the isotropy condition before constructing the corresponding statistical manifold.
\section{Isotropic oscillator}
\begin{figure}
	\includegraphics[width=8cm]{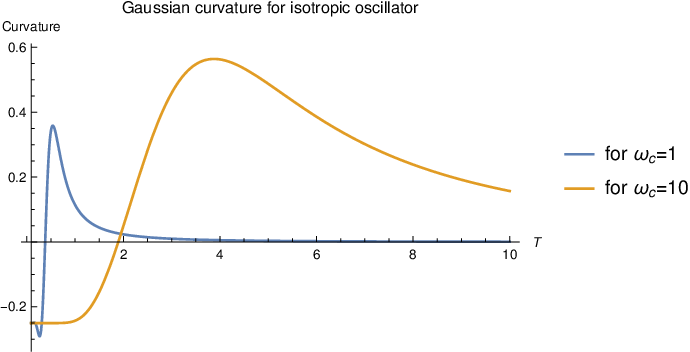}
	\caption{ The variation of the Gaussian curvature $K_{G,iso}$ for   an isotropic oscillator. The curves correspond to two values of the cyclotron frequency $\omega_{c}=1,10$. }
	\label{Piso}
\end{figure}

For an isotropic oscillator, we take $\omega_x=\omega_y=1$ (i.e., $\delta=1$) and $m=1$. The normal mode frequencies are $\Omega_{1,2}=\omega\pm\omega_c/2$, where $\omega=\sqrt{1+\omega_c^2/4}$. 	The covariance matrix of the thermal state is given by
\begin{equation}
	\Sigma_{I}=\frac{\hbar}{2}\left(\begin{array}{cc}
a\mathbb{I}_2 & c_{\mu}J_2\\
-c_{\mu}J_2 & b\mathbb{I}_2
	\end{array}\right),
\mbox{with}\; a=\frac{\mu_1+\mu_2}{2\omega},\;
b=\frac{\omega}{2}(\mu_1+\mu_2),\;
c_{\mu}=\frac{\mu_2-\mu_1}{2}.
\end{equation}
Accordingly, the two-parameter family $\vartheta=(\beta,\omega_c)$ provides the metric tensor components for an isotropic oscillator as
\begin{eqnarray}
	g_{\beta\beta}=\hbar^2(\chi_1+\chi_2),\;
	g_{\omega_c\omega_c}= \frac{\hbar^2\beta^2}{4\omega^2}(\chi_1+\chi_2)+\frac{\omega_c^2(\mu_1+\mu_2)^2}{32\omega^4\mu_1\mu_2}
	\\
	 g_{\beta\omega_c}= \frac{\hbar^2\beta}{2\omega}(\chi_1-\chi_2),\;
	 \mbox{with}\; \chi_j=\Omega_j^2 \cosech^2(\beta\hbar\Omega_j);\; j=1,2.
\end{eqnarray}
Using the metric tensor components, we can evaluate the Gaussian curvature, which can be written as
\begin{equation}
	K_{G,iso}=\frac{\kappa_0\kappa_1^2\kappa_2^2}{\eta_2^2} \left[(\Delta_1-\Delta_2)\eta_1-2\kappa_0(1-\nu^2)\eta_{-1} -\frac{\kappa_0}{4}\left(\frac{1}{\kappa_2}- \frac{1}{\kappa_1}\right)^2\eta_2\right],
\end{equation}
where we have introduced the following quantities
\begin{eqnarray}
	\kappa_0 &=&\frac{\omega_c}{2\omega}, \;
	\nu=\frac{\sinh(\hbar\beta\omega_c/2)}{\sinh(\hbar\beta\omega)},\; \kappa_j=\frac{\sinh\nu_{0j}}{\nu_{0j}},\;
	 \nu_{0j}=\hbar\beta\Omega_j,\\
	 \Delta_j &=& \nu_{0j}\coth\nu_{0j}-1; j=1,2.\;\;
	 \eta_{-1} = \Delta_1\Delta_2 -\nu\left(\frac{\Delta_1}{\kappa_2}-\frac{\Delta_2}{\kappa_1}\right) +\frac{\nu^2}{4} \left(\frac{1}{\kappa_1}-\frac{1}{\kappa_2}\right)^2.\\
	\eta_{1} &=& (1-\nu^2)(1-\kappa_0^2)+\nu\kappa_0^3(\kappa_1+\kappa_2)/2,\;
	\eta_{2} = 2(1-\nu^2)+\kappa_0^2 (\kappa_1^2+\kappa_2^2).	
\end{eqnarray}
The temperature dependence of $K_{G,iso}$ is illustrated in the figure FIG.~\ref{Piso}.
In the zero-temperature limit, the curvature converges to
\begin{equation}
	\lim\limits_{\beta\to\infty}K_{G,iso}=-1/4,
\end{equation}
consistent with the limiting value in Eq.~\eqref{lim1}. In contrast, the Gaussian curvature vanishes along the zero-field locus at every finite temperature $\left(\lim\limits_{\omega_c\to 0}K_{G,iso}=0\right)$. Consequently, the zero-temperature and zero-field limits of the Gaussian curvature do not commute.
Note that at absolute zero, the system is in its ground-state. Consequently, the ground-state covariance matrix is parametrized solely by  $\vartheta=\omega_c$, and the corresponding FR-line element reads
\begin{equation}
	ds_{FR,0}^2=\frac{2\omega_c^2}{(4+\omega_c^2)^2}d\omega_c^2.
\end{equation}
One can distinguish two ground-states corresponding  to different magnetic-field strengths (at $T=0$), characterized by their respective cyclotron frequencies ($\omega_{c,1}$ and $\omega_{c2}$), through the FR distance between two corresponding covariances matrices $\Sigma_1(\omega_{c,1})$ and $\Sigma_2(\omega_{c,2})$ as
\begin{equation}
	d_{FR,0}(\Sigma_1(\omega_{c,1}),\Sigma_2(\omega_{c,2})) = \frac{1}{\sqrt{2}} \left\vert \ln \left( \frac{4+\omega_{c,2}^2}{4+\omega_{c,1}^2}\right) \right\vert.
\end{equation}
\section{PPT separability}
\begin{figure}
	\includegraphics[width=8cm]{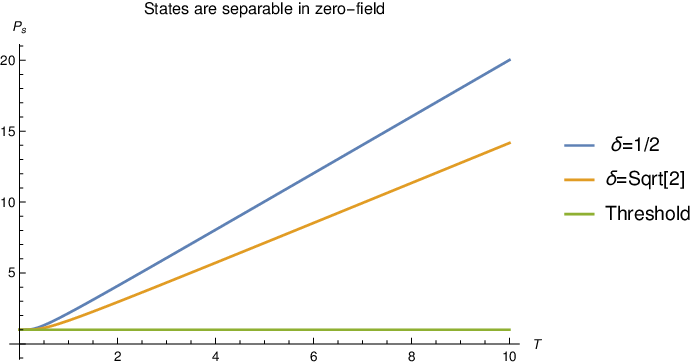}
	\caption{In the absence of an external magnetic field $P_s>1$ at finite temperatures T. The curves correspond to two values of the anisotropy parameter $\delta=1/2,\sqrt{2}$. }
	\label{sepplot1}
\end{figure}

Consider the canonical bipartition of the system with two subsystems $A$ and $B$ as
\begin{equation}
	\xi_A= (\hat{x},\hat{p}_x)^T,\; \xi_B=(\hat{y},\hat{p}_y)^T,
\end{equation}
then the partial transposition (PT) $\Lambda_B$ with respect to subsystem $B$ corresponds to momentum reversal of $B$, while keeping all other coordinates unchanged. In particular,
\begin{equation}
	\Lambda_B=\text{diag}(1,1,1,-1): (\hat{x},\hat{y},\hat{p}_x,\hat{p}_y)\mapsto (\hat{x},\hat{y},\hat{p}_x,-\hat{p}_y).
\end{equation}
Since each subsystem consists of a single bosonic mode (Eq.~\eqref{HN}), the Gaussian positive partial transpose (PPT) criterion is both necessary and sufficient for separability of states. $\Lambda_B$ transforms the covariance matrix $\Sigma$ (Eq.~\eqref{sigma}) as
\begin{equation}
\Lambda_B:\Sigma\mapsto\tilde{\Sigma}=\Lambda_B\Sigma\Lambda_B^T.
\end{equation}
$\tilde{\Sigma}$ has symplectic eigenvalues
\begin{equation}\label{nupm}
	\tilde{\nu}_{\pm}=\frac{\hbar}{2} \sqrt{\frac{1}{2}(\tilde{\Delta}\pm\sqrt{\tilde{\Delta}^2 - 4\mu_1^2\mu_2^2})},\;\; \text{with}\; \tilde{\Delta}=a_1b_1+a_2b_2+2c_1c_2.
\end{equation}
Using Eq.~\eqref{a1}-\eqref{c2} in Eq.~\eqref{nupm}, we write the explicit form of $\tilde{\Delta}$:
\begin{eqnarray}
\tilde{\Delta} &=& \gamma_0(\mu_1-\mu_2)^2+(2+\gamma_1)\mu_1\mu_2;\\
\text{where}\; 
\gamma_0 &=& \frac{(1-\delta^2)^2}{(1-\delta^2)^2+2\omega_c^2(1+\delta^2)+\omega_c^4},\;
\gamma_1= \frac{\omega_c^2(1-\delta)^2}{\delta[(1+\delta)^2 +\omega_c^2]}.
\end{eqnarray} 
Note that $\tilde{\nu}_{+}\ge \tilde{\nu}_{-}$. In order to $\tilde{\Sigma}$ to be a bonafide covariance matrix, it has to satisfy the Peres-Horodecki separability criterion 
\begin{equation}
	\tilde{\Sigma}+\frac{i\hbar}{2}J_4\ge 0,
\end{equation}
which is equivalent to the condition that the smallest symplectic eigenvalue of $\tilde{\Sigma}$ must satisfy
\begin{equation}\label{mathcalP}
	\tilde{\nu}_{-}\ge \hbar/2 \equiv P_s=2\tilde{\nu}_{-}/\hbar \ge 1.
\end{equation}
A bipartite Gaussian state is separable if $P_s>1$, entangled for $P_s<1$, and lies on the PPT threshold for $P_s=1$. Using explicit form from Eq.~\eqref{nupm}, we note that $P_s(T=0,\omega_c=0)=1$, as expected. Moreover, at absolute zero, the state (ground state) is separable for $\omega_c=0$ or $\delta=1$. In other words, both magnetic coupling and anisotropy are necessary for ground-state entanglement. This can be viewed from FIG.~\ref{sepplot1}, in which  we see that the states are always separable ($P_s>1$) for all finite temperature in the absence of $B$.
\begin{figure}
	\includegraphics[width=8cm]{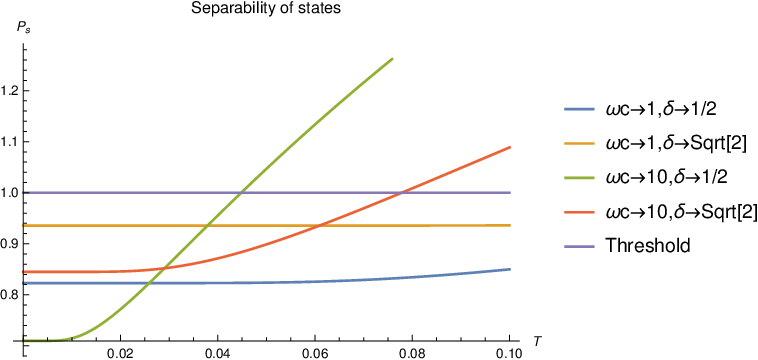}
	\caption{Separability of states requires $P_s>1$. The curves correspond to different  parameter values $(\omega_c,\delta)=\{(1,1/2),(1,\sqrt{2}),(10,1/2),(10,\sqrt{2})\}$. In high temperature limit, all states are separable. Thus, genuine entanglement requires low temperature limit.}
	\label{entplot1}
\end{figure}
From FIG.~\ref{entplot1}, we see that at high temperature limit, all states are separable. This reflects the fact that system tends to show classical limit at high temperature. Genuine quantum effects (entanglement) is manifested at low temperature. Moreover,  both external magnetic field and the anisotropy in oscillator parameters are necessary to induce entanglement in thermal state. FIG.~\ref{entplot1} illustrates the existence of a critical temperature $\beta_{\text{crit}}$ for the entangement threshold. In particular, using ~\eqref{nupm} and ~\eqref{mathcalP}, we see that the critical line is given by
\begin{equation}
	P_{\text{crit}}(\beta_{\text{crit}}, \delta, \omega_c) = \frac{\omega_c^2 (\tau_{+}^2 - \tau_{-}^2)}{4\delta}=1, \text{with}\; \tau_{\pm}= \frac{(1 \mp \delta)^2}{w_{\pm}^2} \sinh^2(\hbar \beta_{\text{crit}} w_{\pm}/2).
\end{equation}
$w_{\pm}$ are given in ~\eqref{covzeroel2}.
\section{Conclusions} 
We have investigated the information geometry and Gaussian separability of a two-dimensional charged anisotropic harmonic oscillator subjected to a homogeneous external magnetic field. Starting from the symplectic diagonalization of the quadratic Hamiltonian, we have obtained the exact thermal covariance matrix and the corresponding Fisher-Rao (FR) metric on the three-parameter statistical manifold characterized by temperature, oscillator anisotropy and cyclotron frequency.\\
In the absence of an external magnetic field, the thermal information manifold admits an isometric embedding into three-dimensional Euclidean space, and its intrinsic Gaussian curvature is found to be strictly negative at finite temperatures. The ground-state manifold exhibits a nonuniform limiting behaviour, wherein the isotropic and zero-field limits of the Gaussian curvature do not commute. The associated curvature singularity originates from the degeneracy of the induced FR metric rather than a singularity of the underlying quantum state. For the isotropic oscillator, the competition between thermal fluctuations and magnetic coupling gives rise to both positively and negatively curved regions. Furthermore, the zero-temperature and zero-field limits of the corresponding Gaussian curvature are shown to be noncommutative.\\
Employing the Peres-Horodecki criterion, we have obtained an exact analytical characterization of Gaussian separability with respect to the canonical Cartesian bipartition. Both oscillator anisotropy and magnetic coupling are necessary and jointly sufficient for ground-state entanglement. At finite temperatures, however, thermal fluctuations suppress entanglement, resulting in a parameter-dependent separability threshold. The isotropic oscillator and the oscillator in the absence of magnetic coupling remain separable at all temperatures.\\
It is worth noting that the intrinsic curvature of the FR manifold does not, by itself, constitute an entanglement witness. Negative curvature, curvature sign reversal and even curvature divergence can occur without the emergence or enhancement of quantum entanglement. This establishes a distinction between the geometric properties associated with statistical distinguishability and the quantum correlations characterized by the PPT criterion.
\section{Data Availability Statement}
No datasets were generated or analysed in the course of this theoretical study.
\section{Acknowledgement}
P. Patra acknowledges financial support from the Anusandhan National Research Foundation (ANRF, formerly SERB), Government of India, under Grant No. EEQ/2023/000784.

\end{document}